\documentclass[aps,prb,amsmath,amssymb,twocolumn]{revtex4-1}
\usepackage{graphicx}
\usepackage{amsmath}
\usepackage{amssymb}
\usepackage[dvipsnames]{xcolor}
\usepackage{indentfirst}
\usepackage{nth}
\usepackage{physics}
\usepackage{bm}
\usepackage[english]{babel}

\newcommand{\des}{\hat{a} }
\newcommand{\cre}{\ensuremath{\hat{a}^{\dagger} } }

\newcommand{\EJ}{\ensuremath{ E_{\text{J}}}}

\newcommand{\im}{\ensuremath{ \text{i} }}

\begin{document}

\title{Josephson Photonics with Simultaneous Resonances}
\author{Kieran Wood,  Andrew D. Armour and Ben Lang}

\affiliation{School of Physics and Astronomy and Centre for the Mathematics and Theoretical Physics of Quantum Non-Equilibrium Systems, University of Nottingham, Nottingham NG7 2RD, UK}
\begin{abstract}
Inelastic Cooper pair tunneling across a voltage-biased Josephson junction in series with one or more microwave cavities can generate photons via resonant processes in which the energy lost by the Cooper pair matches that of the photon(s) produced. We generalise previous theoretical treatments of such systems to analyse cases where two or more different photon generation processes are resonant simultaneously. We also explore in detail a specific case where generation of a single photon in one cavity mode is simultaneously resonant with the generation of two photons in a second mode. We find that the coexistence of the two resonances leads to effective couplings between the modes which in turn generate entanglement.
\end{abstract}
\maketitle

\section{Introduction}

Circuits in which voltage biased Josephson junctions (JJ) are combined with microwave cavities provide an ideal platform for exploring a wide range of microwave photonics. All of the voltage energy associated with tunneling Cooper pairs must be transferred into photons and the properties of JJ-cavity systems can be tuned over a wide range either in-situ or by design,\cite{Hofheinz_2011,Chen_2014,Westig_2017,Rolland_2019,Grimm_2019, Peugeot_2020}. Furthermore, the energy transferred by tunneling Cooper pairs into microwave modes can be tracked by monitoring either the resulting dc current or the microwaves leaking out of the circuit\,\cite{Hofheinz_2011}.  Recent experimental,\cite{Hofheinz_2011,Chen_2014,Westig_2017,Rolland_2019,Grimm_2019, Peugeot_2020} and theoretical work\,\cite{Padurariu_2012,Leppakangas_2013,Armour_2013,Gramich_2013,Dambach_2015,Trif_2015,Souquet_2016,Armour_2017,Leppakangas_2018,Arndt_2019,Kubala_2020,Lang_2021} has explored a wide range of ways in which JJ-cavity systems can be used to generate non-classical microwave states.

Energy exchange between charge carriers and microwaves in JJ-cavity systems is concentrated at resonances where the energy lost by a given Cooper-pair is commensurate with that of the photons in one or more microwave mode(s)\,\cite{Hofheinz_2011,Chen_2014,Westig_2017}. Such resonances can be selected by simply tuning the voltage and are modelled theoretically using a rotating wave approximation (RWA) which leads to a convenient time-independent Hamiltonian for the system\,\cite{Armour_2013,Gramich_2013,Armour_2015,Trif_2015}.  The simplest resonances involve a single mode and can be exploited to provide a single photon source\,\cite{Grimm_2019,Rolland_2019}, although higher order resonances in which two or more photons are generated within a particular mode also occur\,\cite{Armour_2013,Kubala_2015,Lang_2021}. 

Resonances involving two modes (realised, e.g., within the same cavity or in two separate cavities in series with the JJ) can be used to produce entangled photons, via processes in which photons in both are generated simultaneously via a single tunneling process\,\cite{Rolland_2019,Peugeot_2020}. The effective coupling between modes generated by the JJ also supports resonances where Cooper pair tunneling is accompanied by an exchange of photons between modes, processes which could be exploited to engineer efficient heat engines\,\cite{Hofer_2016}.

Despite the very wide range of possibilities offered by JJ-cavity systems, so far attention has generally focused only on cases where a single photon generation/exchange process is resonant. In this paper we instead consider situations where two or more distinct resonant processes can occur at the same time,
leading naturally to competition between them. Here we show how the theoretical formalism used to obtain time-independent Hamiltonians for single-resonance problems can be generalised to address cases with multiple co-existing resonances. We introduce a compact analytic description of the resulting RWA Hamiltonians and show that it leads naturally to an efficient description of the system's classical dynamics. 

\begin{figure}[t]
\centering
{\includegraphics[scale=0.35]{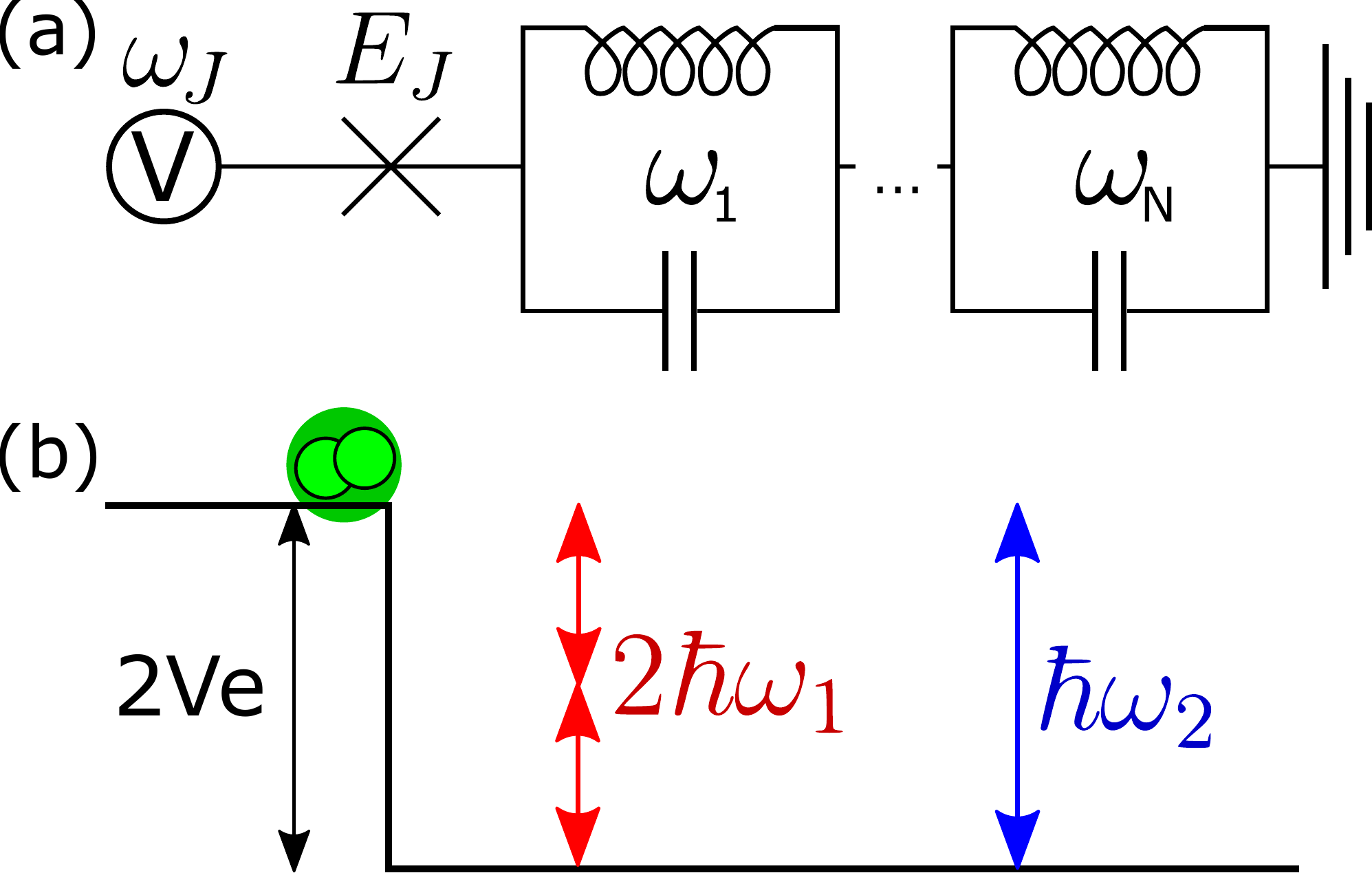}}
\caption{(a) Equivalent circuit of a Josephson junction, biased by a voltage $V=\hbar\omega_J/2e$, in series with a set of $N$ microwave modes modelled as series of LC oscillators with frequencies $\omega_1$, $\omega_2$,\dots, $\omega_N$. (b) As an example we consider the case where a tunneling Cooper pair can transfer energy into photons via two distinct resonant processes: two photons into mode 1 or one photon into mode 2 ($2\hbar\omega_1=\hbar\omega_2=2eV$).}
\label{circuit}
\end{figure}
We illustrate our analysis by investigating in detail a specific example of competing resonances: a two mode system where a single tunneling Cooper pair can generate either two photons in the first mode or one photon in the second mode (see Fig.\ \ref{circuit}). We find that the quantum dynamics doesn't produce a clear `winner' in the competition between resonant processes, instead they can coexist with similar strengths. Furthermore, the coexistence of the resonances generates effective couplings between the modes which can lead to significant entanglement. 

The rest of this article is organised as follows. We start by introducing the theoretical model for the JJ-cavity system in Sec.\ \ref{sec:Model}. In Sec.\ \ref{sec:SF}
we show how special functions can be used to obtain compact expressions for RWA Hamiltonians describing competing resonances and the corresponding classical description is derived in Sec.\ \ref{sec:CSA}. Then in Sec. \ref{sec:TMC} we explore the quantum dynamics that arises for the example with two co-existing resonances. Finally, we conclude in Sec.\ \ref{sec:Con}.

\section{Model System} 
\label{sec:Model}
We consider a system of $N$ harmonic modes, with individual frequencies $\omega_1, \dots, \omega_N$, in series with a JJ and with a voltage bias $V$ applied, as sketched in Fig.\ref{circuit}(a). The modes could be different harmonics within one or more microwave cavities\,\cite{Hofheinz_2011,Chen_2014,Cassidy_2017, Westig_2017,Peugeot_2020}, or they could be realised as lumped element LC-oscillators\,\cite{Rolland_2019}.   
The circuit can be described by the following time-dependent Hamiltonian\,\cite{Armour_2013}
\begin{equation}
\hat{H}= \sum_{n=1}^N  \hbar \omega_{n} \cre_{n} \des_{n} - \EJ \text{cos} \left[ \omega_{\text{J}} t + \sum_{n=1}^N \Delta_{n} ( \cre_{n} + \des_{n} ) \right],
\label{time_dependent_hamiltonian}
\end{equation}
where $\omega_J=2eV/\hbar$, $\des_{n}$ are the annihilation operators of the modes, $\Delta_{n}$ the zero-point displacement (determined by the corresponding mode capacitance, $C_n$, and inductance, $L_n$)  $\Delta_n = (2 e^2 \sqrt{L_l/C_l}/\hbar )^{1/2}$ and $\EJ$ is the Josephson energy of the junction. Almost all of the parameters in this circuit can be varied, either through circuit design\,\cite{Rolland_2019}  ($\omega_n$, $\Delta_n$), or in-situ within a given device e.g.\ via a change of voltage\,\cite{Hofheinz_2011} ($\omega_J$). The value of $\EJ$ can be tuned in-situ by using a parallel combination of two JJs (SQUID) and applying a flux bias.

The time-dependence makes Eq.\ (\ref{time_dependent_hamiltonian}) a difficult Hamiltonian to work with.  In cases where only a single mode is included, resonances where $\omega_J$ is an integer multiple of the mode frequency can be described by an approximate time-independent Hamiltonian obtained via a rotating wave approximation\,\cite{Armour_2013,Gramich_2013,Rolland_2019}. A similar method was applied to study two-mode systems with $\omega_J$ chosen to match the sum of the mode frequencies, defining a single resonance\,\cite{Armour_2015,Trif_2015,Westig_2017}. We will now consider how this approach can be generalised to problems involving a wider set of modes and allowing for cases where more than one process can be resonant. 

Multiple resonances involving a set of $N$ modes arise naturally when their frequencies, and that of the drive frequency $\omega_J$, are all commensurate. 
For convenience, we shall start by assuming that all of the frequencies can be expressed as integer multiples of the fundamental (lowest) mode frequency $\omega_1$: i.e.\ the values of $q_l=\omega_l/\omega_1$ with $l=1,\dots,N$ and $p=\omega_J/\omega_1$ are all positive integers. Resonances in the system associated with the inelastic tunneling of a Cooper-pair across the junction\footnote{Note that we only consider processes involving individual Cooper pairs here\,\cite{Morley_2019}.} are then described by vectors $\bf{m}$, with $N$ integer components that satisfy  $\sum_{l=1}^N q_l m_l=p$, with positive (negative) components $m_l$ describing the gain (loss) of  $|m_l|$ photons in the $l$-th mode. In cases where more than one such vector can be found the system has competing resonances.  

For the simple competing resonance illustrated in Fig.\ \ref{circuit}b we have $N=2$ and $\omega_J=\omega_2=2\omega_1$, hence $p=2$ and the set $\{q\}=(q_1,q_2)=(1,2)$. We can think of this as a competition between two resonances, as to lowest order in the number of photons created/destroyed, creation of either one photon in mode 2 or two photons in mode 1 are both resonant. However, the behavior described by Eq.\ (\ref{time_dependent_hamiltonian}) is rather more complex, and higher order processes involving an exchange between the modes must also be accounted for. In fact, all vectors of the form  ${\bf{m}}^{(k)} = (2k,1-k)$ satisfy the resonance condition with $k=0,\pm 1,\pm 2,\dots$. This illustrates the basic problem in dealing with competing resonances: as soon as there are two modes with frequencies that are both commensurate with $\omega_J$, direct processes in which just one mode, or the other, is excited by inelastic tunneling are accompanied by a whole host of others in which photons are exchanged between the modes. This is a manifestation of the complex mode-mode coupling that the Hamiltonian (\ref{time_dependent_hamiltonian}) gives rise to.

In the following we will consider systems where the resonance condition(s) are met up to some small detunings, $\delta_l$, such that $\omega_l = (q_l/p) \omega_J + \delta_l$ (with $q_l,p$ positive integers and $q_1=1$ as before). We proceed by transforming into a rotating frame via the unitary transform:
\begin{equation}
\hat{U}(t) = \exp(\im \sum_{l=1}^{N} (q_{l}/p) \omega_J  \hspace{0.2pc} \cre_{l} \des_{l} \hspace{0.2pc} t).
\label{unitary}
\end{equation}
The RWA is then made, assuming that terms that retain a time-dependence in the rotating frame can be neglected. This is equivalent to assuming that only the terms describing (close to) resonant processes need to be retained.

The simplest way of expressing the resulting Hamiltonian is to simply pick out the matrix elements in the number state basis that have no time dependence in the rotating frame\,\cite{Dambach_2015}. For the multi-mode case we can do this formally via a filter 
which selects only the relevant time-independent terms. This results in the following recipe for the RWA Hamiltonian 
\begin{eqnarray}
\hat{H}_{\text{RWA}} & =& \sum_{n=1}^{N} \hbar \delta_{n} \cre_{n} \des_{n}\nonumber\\
&&- \frac{\EJ}{2}  \left\{ \mathcal{E} \left[ \text{e}^{ \im \sum_{n=1}^{N} \Delta_{n} \left( \cre_{n} + \des_{n} \right)}  \right] + \text{h.c.} \right\},
\label{Filter_RWA}
\end{eqnarray}
with the filter, $\mathcal{E}$, defined by the relation
\begin{equation}
\mathcal{E}[ \hat{\mathcal{O}} ] = \sum_{\bf{n}} \sum_{{\bf{m}}\in \bf{S}} \ketbra{\bf{n}}{\bf{n}} \hat{\mathcal{O}} \ketbra{\bf{n}+\bf{m}}{\bf{n}+\bf{m}},
\label{filter}
\end{equation}
where $|{\bf{n}}\rangle=|n_1,n_2,\dots,n_N\rangle$ is an $N$-mode Fock state. The sum over $\bf{n}$ runs over all states whilst the other sum is over the vectors ${\bf{m}}$ belonging to the set $\bf{S}$ that satisfy the resonance constraint, $\sum_lq_lm_l=p$, whilst also having $n_l+m_l\ge 0$ for all $l$. Hence for the 2-mode competing resonance where $\omega_J=\omega_2=2\omega_1$, the set $\bf{S}$ is over the vectors ${\bf{m}}^{(k)} = (2k,1-k)$,  leading to the states $|n_1+2k,n_2+1-k\rangle$, with $k$ an integer within the range $-n_1/2\leq k \leq n_2+1$.

In addition to the coherent drive represented by Eq.\ (\ref{Filter_RWA}) a model of the system dynamics must also include the inevitable photon leakage from the modes. This could represent unwanted losses, coupling to collection lines or a mixture of the two. For simplicity we assume a standard zero-temperature Lindblad master equation\,\cite{Carmichael_book}
\begin{equation} 
\dot{\rho} = -\frac{\im}{\hbar}[H_{\text{RWA}}, \rho] + \sum_{l} \frac{\gamma_{l}}{2} ( 2 \des_{l} \rho \cre_{l}  - \cre_{l} \des_{l} \rho - \rho \cre_{l} \des_{l} ),
\label{Master_equation}
\end{equation}
with $\gamma_l$ the loss rate for mode $l$. 

\section{Special Function Form of Hamiltonian}
\label{sec:SF}
The filtering out of the resonant terms to produce a power series embodied by Eq.\ (\ref{Filter_RWA}) is a convenient route for numerical calculations, but  it is difficult to connect with simpler approximate descriptions based, e.g.\ on a coherent state ansatz (see Sec.\ \ref{sec:CSA} below) in particular. Instead it is convenient to derive compact functional forms for the power series of operator terms left after the RWA has been implemented, an approach which is facilitated by the use of normal ordering. 

In the simple case of a single mode system where $\omega_1 = (\omega_J/p) +\delta$,
the Taylor series of a Bessel function can be identified in the normally-ordered expansion that follows after the RWA is made. This leads to the compact expression\,\cite{Armour_2013,Gramich_2013}
\begin{eqnarray}
\hat{H}^{(1)}_{\text{RWA}}& =& \hbar \delta \cre \des \nonumber\\
&& -\frac{\tilde{\EJ}}{2}  : \left( \frac{ (\im \des)^p + (-\im \cre)^p}{(\cre \des)^{p/2}} \right) J_p(2\Delta \sqrt{\cre \des }):,
\label{1_mode_Ham}
\end{eqnarray}
where $:.:$ indicates normal order, $J_p(x)$ is a Bessel function of the first kind of order $p$ and $\tilde{\EJ}=\EJ{\rm{e}^{-\Delta^2/2}}$ is the renormalised value of the Josephson energy\,\cite{Rolland_2019,Leger_2019}. For single-resonance circuits containing multiple cavities, the normally ordered operator power series in the RWA Hamiltonians can be written as products of Bessel functions, one for each cavity involved in the process\,\cite{Armour_2015,Trif_2015}. 

We now generalise this approach to find a compact representation for the RWA Hamiltonian for situations where two or more resonances compete. To do so we go back to consider the full Hamiltonian [Eq.\ (\ref{time_dependent_hamiltonian})] transformed into the rotating frame using Eq.(\ref{unitary}). Making the RWA by discarding the time dependent terms leads to complicated algebra which significantly complicates an analytic derivation, but this difficulty is sidestepped by instead introducing the RWA with an integral over the fundamental period, $T  = 2p\pi/\omega_J$,
\begin{eqnarray}
&&\hat{H}_{\text{RWA}} = \sum_{l=1}^{N} \hbar \delta_{l} \cre_{l} \des_{l} \label{Heff}\\
&&-\EJ \int_{-T/2}^{+T/2} \,\frac{dt}{T} \text{cos} \left[ \omega_J t 
+\sum_{l=1}^{N} \Delta_{l} \left( \cre_{l} e^{\im (q_l/p) \omega_J t} + {\rm{h.c.}} \right) \right]
\nonumber.
\end{eqnarray}
The time dependent terms average to zero over a period, so this integral form is fully equivalent to directly discarding those terms. Formulating the RWA Hamiltonian in this form provides a straightforward way to express it in terms of special functions which can be defined via integrals as we now show.

To simplify Eq.\ (\ref{Heff}) for the $N$ mode system, we introduce special functions denoted $Z$, defined via the generating function
\begin{equation}
\sum_{n=-\infty}^{\infty} Z_{p}^{\{{q}\}}({\bf{\hat{x}}}) y^n = :\exp\left[\frac12 \sum_{l=1}^{N} \left( \hat{x}_{l} y^{q_{l}} - \frac{\hat{x}_{l}^\dagger}{y^{q_{l}}} \right) \right] :,
\label{generator}
\end{equation}
with the colons indicating normal ordering, as usual. The function $Z_{p}^{\{{q}\}}({\bf{\hat{x}}}) = Z_{p}^{q_1, q_2, ... q_N}(\hat{x}_1, \hat{x}_2,... \hat{x}_N)$ with $\hat{x}_l$ in our case a mode raising or lowering operator, up to a constant factor. The $N$ superscript indices, $\{q\}$, together with the subscript index, $p$, together fully encode the resonance conditions that will need to be incorporated in the reformulation of Eq.\ (\ref{Heff}). These functions are essentially multi-dimensional Bessel functions\,\cite{Dattoli1998, Korsch2008}, but with minor modifications to incorporate complex and operator arguments more readily. As with the single mode case, normal ordering removes all ambiguity from the corresponding power series involving operator arguments.

For a single mode ($N=1$) case the $Z$ function for the c-number argument $A{\rm{e}^{\im \theta}}$ is just an ordinary Bessel function multiplied by a phase factor, $Z^{(1)}_{p}( A {\rm{e}}^ {\im \theta} ) = J_p(A) {\rm{e}}^{\im p\theta}$. The two mode ($N=2$) version of the $Z$ function is closely related to the 2D generalisation of the Bessel function\,\cite{Korsch2008}. Many properties of these functions, such as Taylor series representations, derivatives and relational properties are derived in Appendix \ref{Appendix:z_functs}. These relations prove to be surprisingly simple, allowing expressions involving the $Z$ functions to be manipulated quite straightforwardly.

A useful integral representation of the $Z$ functions is obtained by setting $y=\exp(\im t)$ in Eq.\ (\ref{generator}), then inserting a factor of $ (1 / 2\pi) \int_{-\pi}^{\pi}dt \exp(-\im m t)$ on both sides of the equality. Noticing that on the left the integral reduces to a Kronecker $\delta$-function\,\cite{Korsch2008}, one finds
\begin{equation}
Z_{p}^{\{ q \}} ({\bf{\hat{x}}}) =  : \int_{-\pi}^\pi \frac{dt}{2\pi} \exp{ \sum_{l=1}^N \frac12 \left(\hat{x}_l {\rm{e}}^{\im q_l t} - {\rm{h.c.}} \right) -\im p t }:.
\label{NDZint}
\end{equation}

Returning to the RWA Hamiltonian, the expression is simplified by splitting the cosine in Eq.\ (\ref{Heff}) into a sum of exponentials, each of which is rearranged to achieve normal order, and then identifying the integral representations of the $Z$ functions, Eq.\ (\ref{NDZint}). The Hamiltonian can therefore be expressed as
\begin{eqnarray}
\hat{H}_{\text{RWA}} &=& \sum_{l} \hbar \delta_{l} \cre_{l} \des_{l}-\frac{\tilde{E}_J}{2} \left[  Z_{p}^{\{ q \}}( {\bf{\hat{x}}} ) + {\rm{h.c}}\right]
\label{HRWA}
\end{eqnarray}
where  $\hat{x}_l=2\im \Delta_l \des_l$ 
and we have redefined $\tilde{E}_J = {\EJ} \exp[-\sum_{l=1}^{N} \Delta_{l}^2 / 2]$. 
Although apparently rather abstract, Eq.\ (\ref{HRWA}) facilitates analytic manipulations, as we demonstrate in the next section. 

As expected, the general expression, Eq.\ (\ref{HRWA}), reduces to Eq.\ (\ref{1_mode_Ham}) in the single mode limit ($N=1$). Similarly the Hamiltonians considered in Refs.\ [\onlinecite{Armour_2015, Trif_2015, Souquet_2016, Dambach_2017_2, Hofer_2016, Dambach_2017}] are recovered for cases with a unique resonance, but more than one mode.



\section{Coherent State Ansatz}
\label{sec:CSA}
A coherent state ansatz can be used to obtain a simpler approximate description of the system's dynamics\,\cite{Armour_2013,Meister_2015,Morley_2019}. The idea is to assume that each mode is in a coherent state, $\rho_{\alpha} = \bigotimes_{l=1}^N \ket{\alpha_{l}}\bra{\alpha_{l}}$, described by a complex amplitude $\alpha_{l}$. Substituting this into the master equation (\ref{Master_equation}) leads to a set of equations of motion for the amplitudes, the fixed points of which provide a valuable framework for understanding the dynamics of the system\,\cite{Armour_2013, Bartolo_2016,Roberts_2020, Lang_2021}. This approach
can be thought of as providing an essentially classical description of the dynamics as (quantum) fluctuations in the amplitudes are neglected\,\footnote{Note, however, that one could in principal extend this approach to include small quantum fluctuations about the classical fixed points\,\cite{Armour_2013,Armour_2017}}. When applied to systems with a unique resonance, the resulting fixed point amplitudes have been shown to provide an increasingly accurate way of predicting properties like the average occupation numbers of the modes as the strength of the quantum fluctuations (measured by $\Delta_l$) are reduced\,\cite{Armour_2017}. However, one key limitation is that no information is provided about the way in which the density operator spreads between two or more coexisting stable fixed points\,\cite{Lang_2021}.

To apply the coherent state ansatz to the competing-resonance case we evaluate $\dot{\alpha_l}= \Tr( \des_l \dot{\rho} )$ using Eq.(\ref{Master_equation}), exploiting 
the relation $[  \des,f(\des,\cre) ] = \partial f(\des,\cre)/\partial \cre$, and then substituting in $\rho_{\alpha}$. The analytic properties of the $Z$ functions (discussed in Appendix \ref{Appendix:z_functs}), and in particular their derivatives [see Eq.\ (\ref{derivs})], make this a straightforward calculation. The amplitudes are thus found to obey the following coupled set of equations:
\begin{eqnarray}
\dot{\alpha}_{l} &=& -\left( \im \delta_{l} + \frac{\gamma_{l}}{2} \right) \alpha_l - \frac{\tilde{E}_J  \Delta_{l}}{2\hbar} \left[ Z_{p + q_{l}}^{\{ q \}} ( \{ 2 \im \Delta_{m} \alpha_{m} \} ) \right.\nonumber\\
&&\left.-  Z_{p - q_{l}}^{\{ q \}} ( \{ - 2 \im \Delta_{m} \alpha^*_{m} \} ) \right].
\label{fixed_point_equ}
\end{eqnarray}

The classical fixed point(s) are obtained by solving the algebraic equations for $\alpha_l$ obtained by setting $\dot{\alpha_l}=0$ for all $l$. The stability of these points (and hence their role in the system's long time dynamics) is determined by the eigenvalues of the Jacobian matrix, 
\begin{equation}
J = \begin{pmatrix} \frac{\partial}{\partial \alpha_1} & \frac{\partial}{\partial \alpha_1^*} & \frac{\partial}{\partial \alpha_2} & \frac{\partial}{\partial \alpha_2^*}... \end{pmatrix}^T \begin{pmatrix} \dot{\alpha}_1 & \dot{\alpha}_1^* & \dot{\alpha}_2 & \dot{\alpha}_2^*... \end{pmatrix} \, .
\end{equation}
If any of the eigenvalues are positive the classical fixed point is unstable.

The fact that the Z-functions can be differentiated and evaluated fairly easily is invaluable in locating the stable fixed points of the classical system. In particular, Eq.\  (\ref{NDZint}) provides a convenient way of carrying out the numerical evaluations of $Z$ functions with complex arguments that arise when calculating the fixed points and the Jacobians needed to determine their stability.

\section{Example: Two Mode Competition}
\label{sec:TMC}
Having obtained formal expressions for the RWA Hamiltonian in cases where competing resonances exist, we now look in detail at the specific example of the two-mode problem with $p=2$ and $\{q\}=(1,2)$ (see Fig.\ \ref{circuit}b). Our main aim is to gain insight into how competing resonances can affect the quantum dynamics of the system, but the analysis also serves to illuminate the very general formulations presented in the preceding sections.

On-resonance, the RWA Hamiltonian (\ref{HRWA}) for our two mode system with competing resonances takes the form
\begin{equation}
H_{\text{RWA}}^{(2)} = -\frac{\tilde{E}_J}{2} \left[Z_{2}^{1,2}( 2 \im \Delta_{1} \des_{1}, 2 \im \Delta_{2} \des_{2} ) + \text{h.c.} \right].
\end{equation}
We note that one can use the properties of the Z-functions detailed in Appendix \ref{Appendix:z_functs} to re-express this as an infinite sum over products of Bessel functions of different orders, or equivalently as an operator power series with three nested summations, but the resulting expressions are unwieldy. 
However, some insight into the interaction between the modes can be gained by analysing the parts of the Hamiltonian related to the lowest order  processes. Including just the terms up to $4^{\text{th}}$ order in the creation/annihilation operators:
\begin{eqnarray}
H_{\text{RWA}}^{(2)} \approx -\frac{\tilde{E}_J}{2} : & \left[  \im \Delta_2 \des_2  \left(1 - \Delta_1^2 \hat{n}_1 - \frac{\Delta_2^2 \hat{n}_2 }{2}\right) \right. \nonumber \\
-& \frac{1}{2}\left(\Delta_1 \des_1\right)^2 \left(1 - \frac{\Delta_1^2 \hat{n}_1}{3} - \Delta_2^2 \hat{n}_2 \right) \nonumber \\
 +&  \left. \frac{1}{4}\left(\Delta_1 \Delta_2 \cre_1 \des_2\right)^2 + \text{h.c.} \right] :. \label{int}
\end{eqnarray}
with $\hat{n}_l = \cre_l \des_l$. We can see that two qualitatively rather different effects are present. The first two lines of (\ref{int}), together with their corresponding Hermitian conjugates,  describe processes where photons are added/removed to just one of the modes, but in both cases the effective rates are modified by the photon populations of \emph{both} modes. This provides a form of nonlinear coupling similar to that arising, for example, in cavity optomechanics\,\cite{Aspelmeyer_2014}. In contrast, the last line of (\ref{int}) consists of a more direct form of interaction involving the conversion of quanta between the modes, though it occurs at $4^{\text{th}}$ order in the operators. 

It is interesting to note that this effective interaction is very different to that which arises for a single resonance where the energy from a Cooper pair generates photons in two modes simultaneously\cite{Armour_2015,Trif_2015,Dambach_2017,Peugeot_2020}. In the latter case, the interaction contains terms which are bilinear, reducing to a degenerate parametric amplifier to lowest order in the operators.

\subsection{Fixed Point Analysis}

\begin{figure}[t]
\centering
\includegraphics[scale=0.7]{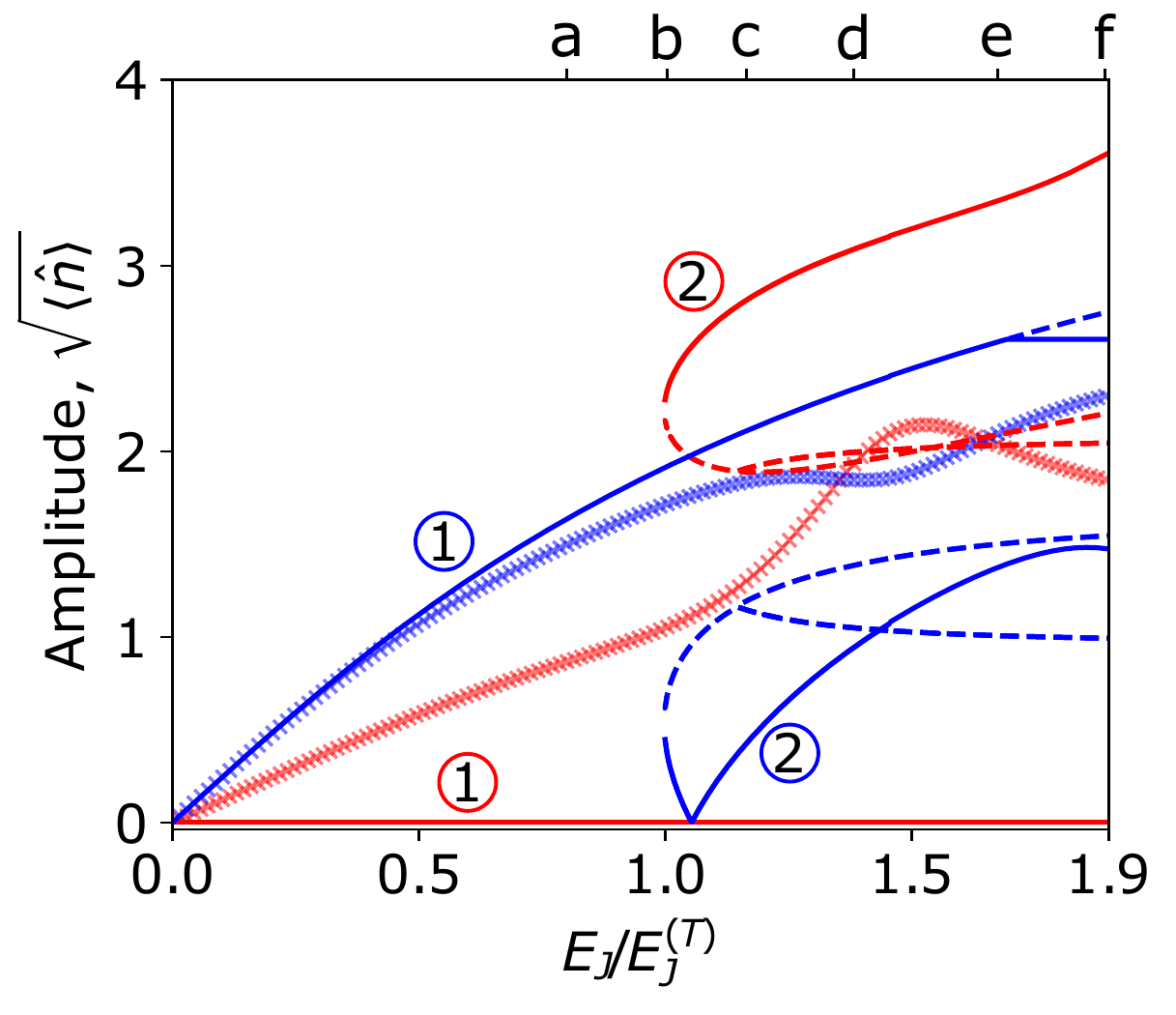}
\caption{Classical fixed point amplitudes, $|\alpha_1|$ (red) and $|\alpha_2|$ (blue) as a function of $E_J/E_J^{(T)}$, 
full (dashed) lines indicate points that are stable (unstable), with $E_J^{(T)}$ the threshold beyond which both modes can be excited. Numbers \textcircled{1},\textcircled{2} indicate the pairings between the amplitudes of the individual modes at the two stable fixed points.
 Also shown for comparison are results from the numerical solution of the master equation (crosses): $\sqrt{\langle \hat{n}_1}\rangle$ (red) and $\sqrt{\langle \hat{n}_2\rangle}$ (blue).  We have set $\Delta_1 = 0.5$, $\Delta_2 = \Delta_1 / \sqrt{2}$ and $\gamma_1=\gamma_2$ unless otherwise indicated. Labels (a)-(f) on the upper axis indicate $E_J / E_J^{(T)}$ values illustrated in Fig.\ \ref{quantum_solutions}.}
\label{Fixed_points}
\end{figure}

\begin{figure*}[t]
\centering
\includegraphics[scale=0.65]{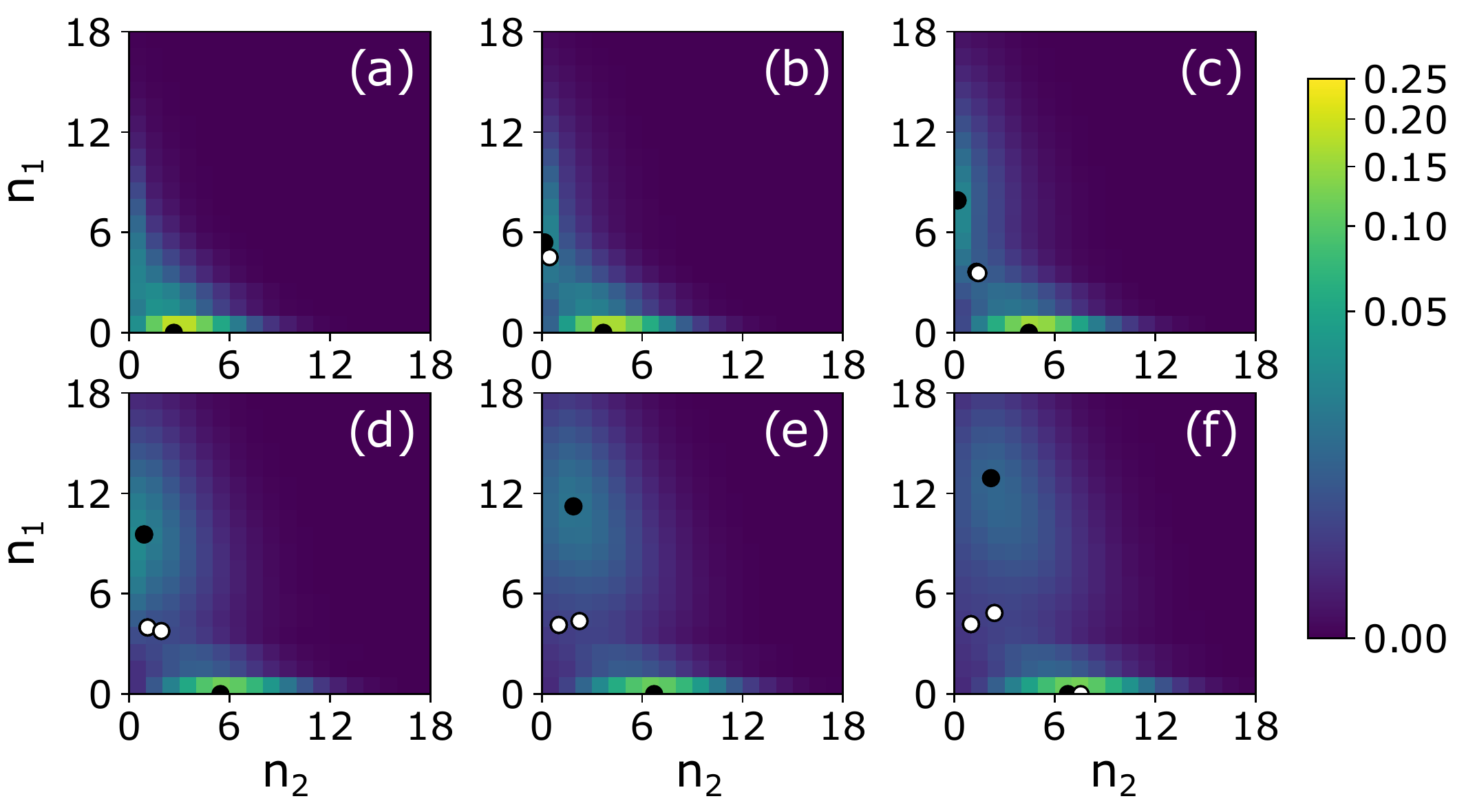}
\caption{Steady state joint photon number distribution, $P(n_1,n_2)$, for a variety of $E_J/ E_J^{(T)}$ values: (a) 0.80 (b) 1.00 (c) 1.16 (d) 1.38 (e) 1.67 and (f) 1.89. In each panel the color of the pixel at location ($x$, $y$) indicates the probability of finding exactly $y$ photons in the first mode and exactly $x$ in the second. The filled (empty) circles superimposed on the number distribution indicate the locations of the stable (unstable) classical fixed points. The probabilites found in the quantum case are concentrated near the stable classical solutions.} 
\label{quantum_solutions}
\end{figure*}
For our two mode case, the equations of motion for the
mode amplitudes that follow from (\ref{fixed_point_equ}) are 
\begin{eqnarray}
\dot{\alpha}_1 &=& -\frac{\gamma_{1}}{2}\alpha_1   - \frac{\tilde{E}_J \Delta_{1}}{2\hbar} \left[ Z_{3}^{1,2} (x_1, x_2 ) -  Z_{1}^{1,2}(x_1^*,x_2^* ) \right] \nonumber \\ 
\dot{\alpha}_2 &=& -\frac{\gamma_{2}}{2} \alpha_2   - \frac{\tilde{E}_J \Delta_{2}}{2\hbar} \left[ Z_{4}^{1,2} (x_1,x_2 ) -  Z_{0}^{1,2}(x_1^*, x_2^* ) \right], \nonumber
\end{eqnarray}
with $x_j=2 \im \Delta_j\alpha_j$. To obtain the corresponding fixed points we use standard optimisation methods, evaluating the $Z$ functions through numerical integration (\ref{NDZint}). It is possible to instead proceed by splitting the $Z$ functions into sums over products of Bessel functions (see Appendix \ref{Appendix:z_functs}). However, direct use of $Z$ functions, evaluated by integration, has a number of advantages. Firstly, it readily scales to higher dimensions (more modes)\footnote{It should be noted that for systems with larger numbers of modes, the necessary optimisation takes longer and the number of fixed points also tends to increase.}. Secondly, it avoids the subtleties of working out where to truncate the (in principle infinite) summations that arise. Indeed, we found the integration method to be much faster in our calculations.

The fixed points are given by pairs of values $\alpha_1,\alpha_2$, the amplitudes of which are shown in Fig.\ \ref{Fixed_points} as a function of the drive strength, $\EJ$, normalised by a threshold value $\EJ^{(T)}$ (defined below). Initially there is only one stable solution with the amplitude of mode 2 (which is resonantly driven $\omega_2=\omega_J$) growing linearly at first whilst the amplitude of mode 1 ($\omega_1=\omega_J/2$) remains zero throughout. This fixed point represents the case where mode 2 wins completely in the competition between resonances. Indeed, the behavior of $\alpha_2$ for this fixed point matches exactly what one gets with a single resonantly excited mode\,\cite{Armour_2013}: it grows more slowly with increasing $E_J$ and its amplitude eventually becomes locked to a constant value (at $E_J /E_J^{(T)}\simeq 1.69$).

At larger drive strengths the picture changes significantly with a second stable fixed point emerging. A saddle-node bifurcation occurs at $\tilde{E}_J/ \hbar \gamma \approx 6.87$, where we have assumed $\gamma=\gamma_1=\gamma_2$. Since the first mode can now become excited, we use this  bifurcation point to define the threshold value for the drive strength, $E_J^{(T)}$.  The bifurcation is collective: the amplitudes of both modes change abruptly. The new stable solution has nonzero amplitudes in both modes, though with that of mode 1 significantly larger than that of mode 2. The threshold occurs at a higher drive strength than that which is required to excite a single mode at the two-photon resonance\,\cite{Armour_2013}, and hence one can think of the presence of the resonantly driven mode 2 as tending to suppress the excitation of mode 1.

There are in fact two bifurcations that occur simultaneously at the threshold, although the two are identical up to phases leading to pairs of fixed points with matching amplitudes, leading to only one set of curves in Fig.\ \ref{Fixed_points}\footnote{The two bifurcations and their fixed points are related by the symmetry expressed in (\ref{Phase_form}) with $\theta=\pi$, recalling that we have set $p=2$.}. Interestingly, the amplitude in mode 2 of the new stable points initially drops with increasing drive, until it touches zero for 
$E_J/ E_J^{(T)} \sim 1.055$, after which it grows again. Seen in the full phase space the complex amplitude of the fixed point moves continuously through the origin. We can think of this second stable fixed point as representing a case where mode 1 wins the competition between resonances, winning completely for $E_J/ E_J^{(T)} \sim 1.055$.


\subsection{Quantum Steady State}
We now move on to examine the full quantum dynamics of the mode competition, using numerical solutions of the master equation (\ref{Master_equation}) obtained using the QuTiP package \cite{QUTIP}. Figure \ref{Fixed_points} compares the steady state expectation values $\sqrt{\langle\hat{n}_1\rangle}$  and $\sqrt{\langle \hat{n}_2\rangle}$ with the stable fixed point amplitudes. Although the connection between these quantities is apparent at low $E_J$ (for mode 2 in particular), it is no longer clear after the bifurcation which leads to bistability with the emergence of the second stable fixed point.

A much clearer understanding of the quantum behavior can be obtained by looking instead at the joint number state probability distribution of the two mode system, given by $P(n_1,n_2)=\langle n_1,n_2|\rho_{ss}|n_1,n_2\rangle$ with $\rho_{ss}$ the steady-state density operator. Several examples of $P(n_1,n_2)$ for different choices of $E_J$ are shown in Fig.\ \ref{quantum_solutions}, overlaid with the locations of the corresponding classical fixed points. Whilst there can be more than one stable classical fixed point for a given parameter set, the quantum dynamics always have a unique  steady state solution. We see that at low $E_J$ the probability distribution is peaked around the location of the only classical stable fixed point, albeit with a significant spread due to quantum fluctuations. For $E_J>E_J^{(T)}$, the probability distribution becomes bimodal with peaks roughly concentrated around the locations of the two co-existing classical stable fixed points. Interestingly, these two peaks have a rather different character: the one corresponding to high occupation of mode 1 (and low occupation of mode 2) is much more diffuse than the one corresponding to high occupation of mode 2 (and low occupation of mode 1). Nevertheless, the overall message is clear: the mode competition has no overall winner in the quantum regime. Instead, both of the classical solutions are represented within the quantum steady state.

\subsection{Mode Correlations}
Finally, we examine the correlations that develop between the two modes that ensue as the quantum system combines the two very different outcomes apparent in the bistability of the fixed points. We will look at amplitude correlations within and between the modes and then quantify the entanglement that is generated.

\begin{figure}
\centering
\includegraphics[scale=0.5]{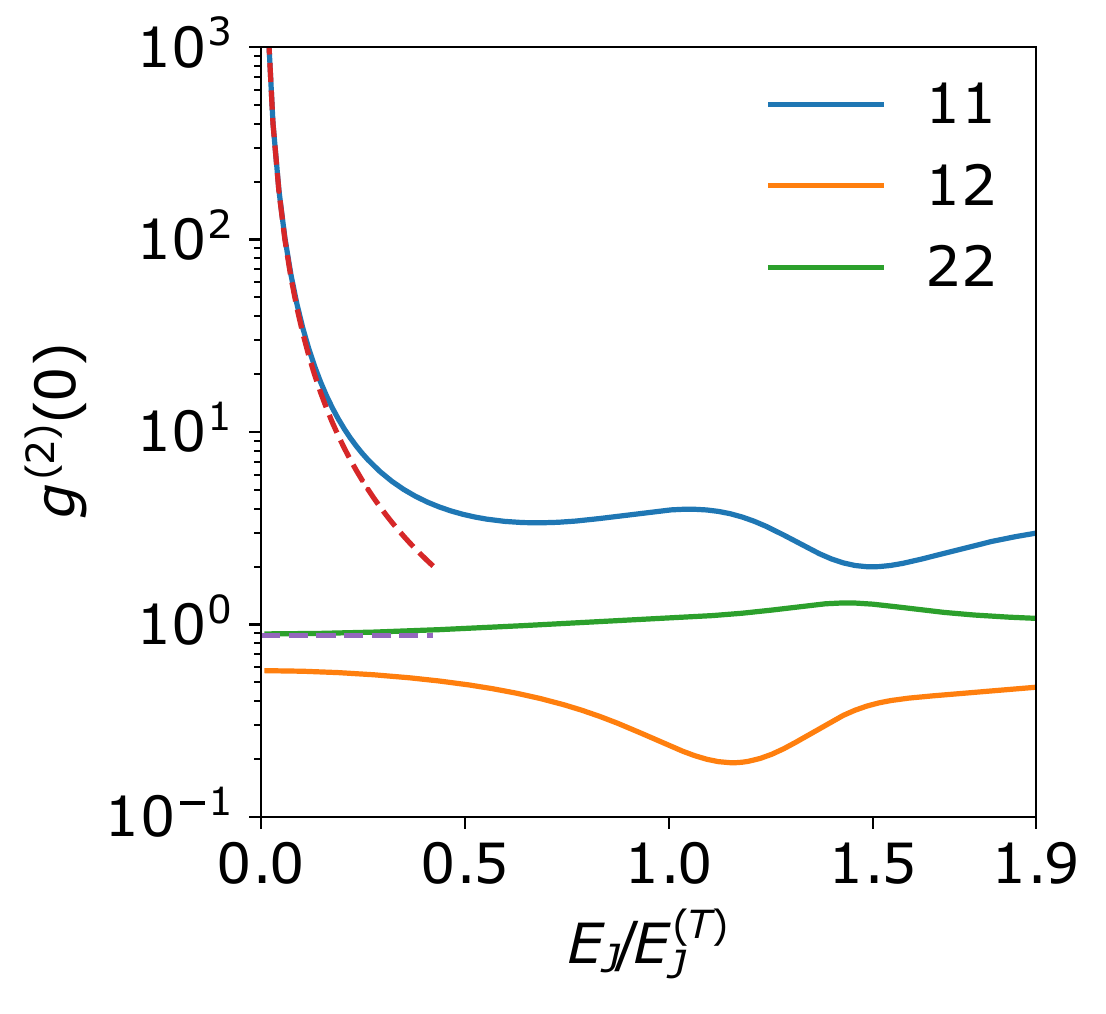}
\caption{Second order correlation functions, $g_{ij}^{(2)}(0)$. Full lines are from numerical calculations, dashed lines are low-$\EJ / \hbar \gamma$ estimates for $g_{11(22)}^{(2)}(0)$ discussed in the text.}
\label{g2s}
\end{figure}

The bimodal number-state distributions that emerge at larger $E_J$ values indicate that the photon populations have become anti-correlated \footnote{Anti-correlation could also be anticipated from the presence of terms that swap quanta between the modes in the interaction Hamiltonian [see Eq.\ \ref{int}].}. 
The detection of a photon from one mode means that it is less likely that one will be found in the other. Such effects can be quantified using second order correlation function\,\cite{Kubala_2015,Westig_2017,Arndt_2019} 
\begin{equation}
g^{(2)}_{ij}(0)=\frac{\langle\hat{a}_i^{\dagger}\hat{a}_j^{\dagger}\hat{a}_i\hat{a}_j\rangle}{\langle \hat{n}_i\rangle \langle \hat{n}_j\rangle}.    
\end{equation}
The auto-correlations for each mode ($i=j=1,2$) and the cross-correlations ($i,j=1,2$) are shown in Fig.\ \ref{g2s}. The auto-correlations are what one would expect for uncoupled modes at low $E_J/\hbar\gamma$, with more complex behavior emerging at larger drive strengths. For mode 1,  photons are always produced in pairs  $(\omega_J=2\omega_1)$ and assuming rare (uncorrelated) pair creation events implies\,\cite{Kubala_2015,Vyas_89} $g_{11}^{(2)}(0) \sim 1/(2 \langle n \rangle)$, which matches the low $\EJ$ behavior very well. For mode 2, photons are produced one-at-a-time ($\omega_J=\omega_2$) and a modest anti-bunching of the photons is expected at low  $\EJ/\hbar\gamma$, taking into account the non-linearity\,\cite{Kubala_2015} $g_{22}^{(2)}(0) \sim (1-\Delta_2^2/2)^2$. In fact $g_{22}^{(2)}(0)$ remains slightly higher than this estimate (at low $\EJ$) and drifts higher still with increasing $\EJ$.
This is a result of coupling to the other mode which opens up the possibility of a range of higher order processes that tend to promote bunching, e.g., one in which inelastic Cooper pair tunneling generates two photons in mode 2 whilst simultaneously annihilating two photons from mode 1.

The cross-correlation, $g_{12}^{(2)}(0)$, remains less than unity throughout indicating the expected anti-correlations. No clear connection to the behavior of the classical fixed points is apparent here, though there is a minimum in $g_{12}^{(2)}(0)$ within the bistable region. Furthermore, the anti-correlation means that the Cauchy-Schwartz inequality $\sqrt{g^{(2)}_{11}(0)g^{(2)}_{22}(0)}\ge g^{(2)}_{12}(0)$  is never violated here. This is in contrast to a single resonance where photons are created in pairs with one in each of two  modes\,\cite{Armour_2015,Trif_2015,Arndt_2019}, thereby generating positive correlations. 

The effective interactions generated by the competing resonances do not just generate anti-correlations in the two modes, they are also able to entangle them even though they are purely nonlinear, involving only terms that are $3^{\rm{rd}}$ order or higher in the creation/annihilation operators [see Eq.\ (\ref{int})]. To demonstrate this we use the log-negativity as a convenient measure of entanglement\,\cite{Dambach_2017,Peugeot_2020}, defined as
\begin{equation}
E_{\mathcal{N}}(\rho)=\log_2\left[1+2{\mathcal{N}}(\rho)\right] , 
\end{equation}
where the negativity, $\mathcal{N}$, is the absolute value of the sum of the negative eigenvalues of the partial transpose of the density operator\,\cite{Vidal_02,Horodecki_2009}. A logarithmic negativity exceeding zero is sufficient (though not a necessary condition) to identify a state as entangled.

\begin{figure}
\centering
\includegraphics[scale=0.5]{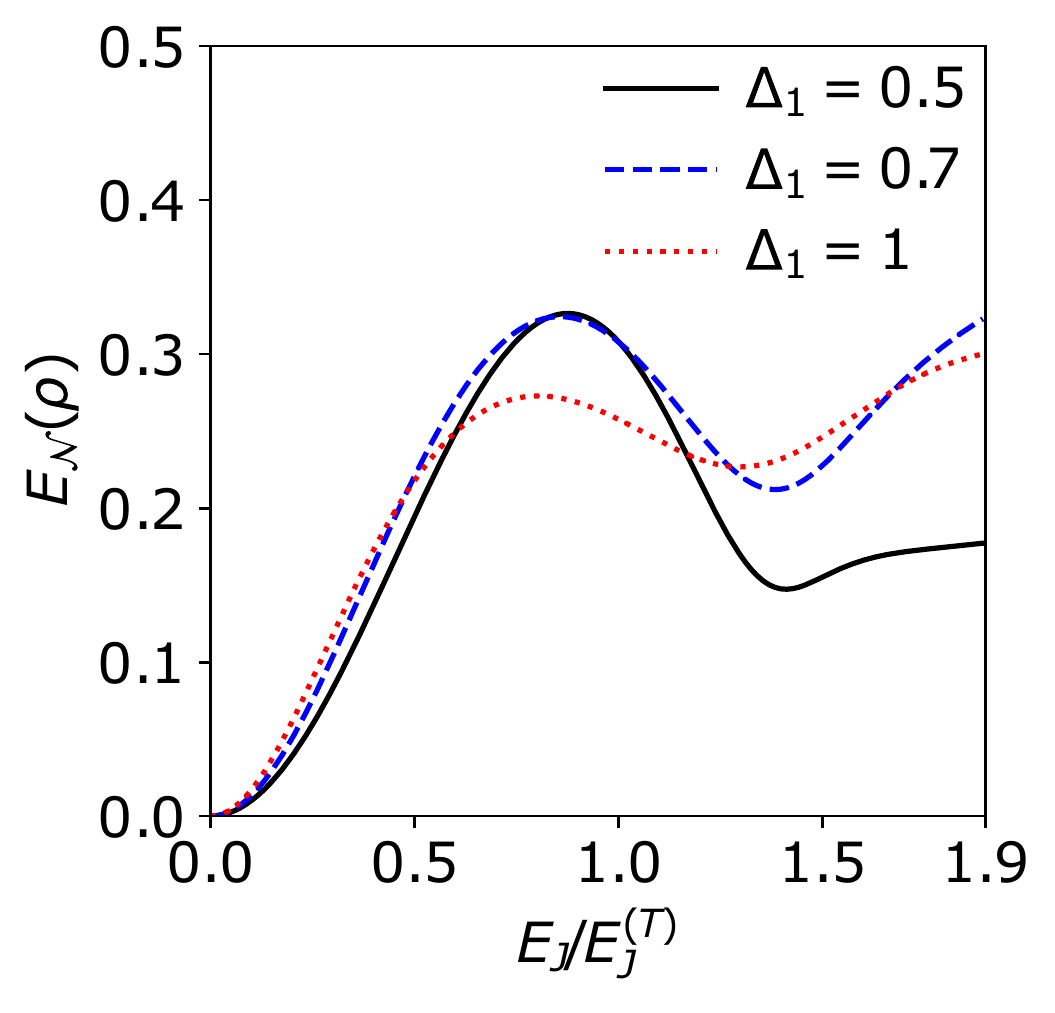}
\caption{Log-negativity in steady state as a function of $E_J$. In each case $\Delta_2 = \Delta_1 / \sqrt{2}$.}
\label{entanglement}
\end{figure}

The behavior of $E_{\mathcal{N}}(\rho)$ as a function of the drive is shown in Fig.\ \ref{entanglement}. We find that the logarithmic negativity initially grows smoothly with the drive strength, later going through a maximum (before the threshold is reached) and then a minimum, but remaining non-zero throughout. The values of the logarithmic negativity achieved are not especially small given the higher order nature of the processes that give rise to the correlations [see Eq.\ \ref{int}]. The peak in Fig.\ \ref{g2s}b is only about a factor of two less than $\ln 2$ which is the upper bound achievable in the two-mode squeezed state produced by a coherent parametric amplifier interaction\,\cite{Woolley_2014}, which is bilinear in the operators.


\section{Conclusions}
\label{sec:Con}
We have explored the quantum dynamics of systems in which inelastic tunneling of Cooper pairs across a voltage biased JJ excites a series of microwave oscillators via two or more competing resonant processes. The competing resonances arise when the mode frequencies and the Josephson frequency (set by the bias voltage) are commensurate. The competition between the resonances can be described by a simplified time-independent Hamiltonian using a RWA, following the approach used for cases with a single resonance. However, the resulting Hamiltonians are rather complicated and unwieldy, even for systems with just two modes. The very strong nonlinearity of the system, together with commensurable mode frequencies, mean that a large number of processes that couple the modes together need to be accounted for. We introduce a compact and efficient technique for analysing such RWA Hamiltonians using normal ordering and a generalised special function. We illustrate the utility of this approach by showing how it can readily be applied to obtain simplified (classical) equations of motion for the amplitudes of the modes. 

We also explored in detail a simple example in which two resonances compete in a two-mode system. Two stable classical fixed points of the system emerge, each one associated with a different one of the two competing resonances clearly `winning'. The quantum dynamics reveals a more complex situation in which bistability emerges naturally with contributions from both resonances evident in the steady-state density matrix. Furthermore, although the effective interactions between the modes in the presence of competing resonances are purely nonlinear, they are sufficient to generate a significant amount of entanglement.

It would be interesting to investigate how competing resonances evolve in cases involving more than two modes in the future. Unfortunately, straightforward numerical solutions of the quantum dynamics become less and less tractable as the state space grows with the number of modes. However, the compact formulations of the multi-mode RWA Hamiltonians developed here should prove a useful starting point for developing analytic approximations. 
\section*{Acknowledgements}
We thank G. Morley for helpful conversations. This work was supported through a Leverhulme Trust Research Project Grant No. RPG-2018-213 and KW received a Summer Student Scholarship funded by the Thomas Farr Charity.

\appendix
\section{$Z$ Functions}
\label{Appendix:z_functs}

The generating function for the $Z$ functions is given by Eq.\ (\ref{generator}). This is very similar to the generating function for multi-dimensional generalisations of the Bessel functions\,\cite{Korsch2008}, consequently $Z$ functions are closely related to Bessel functions ($J_p(x)$):
\begin{eqnarray} 
Z_p^{(1)}(\hat{x}) = Z_p (\hat{x}) &=&  : \sum_{m=0}^{\infty} \frac{(-1)^m (\hat{x}^\dagger \hat{x})^m \hat{x}^p}{m! \, \Gamma(m+p+1) \, 2^{2m+p}} :,
\nonumber\\
Z_p (\hat{x}) &=& :\left( \frac{\hat{x}}{\sqrt{\hat{x}^\dagger \hat{x}}} \right)^p J_p (\sqrt{\hat{x}^\dagger \hat{x}}):. \label{Kdef2} 
\end{eqnarray}
With $\Gamma$ the Gamma function. Note that we will suppress the single superscript for 1D $Z$ functions for brevity. 

Equation (\ref{Kdef2}) indicates that the $Z$-functions have an amplitude set by a Bessel function, but with a different phase, something which becomes immediately apparent if one evaluates the expectation value with a coherent state. Another consequence is that $p$ denotes the overall surplus of powers of $\hat{x}$ over powers of $\hat{x}^\dagger$ in the expression, with negative $p$ naturally indicating a surplus of $\hat{x}^\dagger$ over $\hat{x}$ instead.

The similarity to Bessel functions continues into higher dimensions with two dimensional $Z$ functions close to the 2D generalisations of Bessel functions given in \cite{Korsch2008}. Specifically 2D $Z$ functions can be defined as a series expansion over 1D functions:
\begin{equation}
Z_p^{q_1,q_2} (\hat{x}_1,\hat{x}_2) = \sum_{\mathbf{m} \in \bf{S}} Z_{m_1} (\hat{x}_1) Z_{m_2} (\hat{x}_2) \, .
\label{2DK1}
\end{equation}

More generally $Z$ functions of any dimensionality can be expressed as an infinite sum over a product of $Z$ functions of one fewer dimension with 1D functions:
\begin{equation}
Z_p^{q_1, ... q_N } \left( \hat{x}_1, ...\, \hat{x}_{N} \right) = \sum_{l = -\infty}^{\infty} Z_{p-q_N l}^{ q_1, ... q_{N-1} } \left( \hat{x}_1, ...\, \hat{x}_{N-1} \right) Z_l (\hat{x}_N) .
\label{NDZsum}
\end{equation}
Alternatively, this can be expressed as
\begin{equation}
\begin{split}
Z_p^{\{ q \}} ( {\bf{\hat{x}}} ) &= \sum_{{\bf{m}}\in \bf{S}} Z_{m_1} ( \hat{x}_1 ) Z_{m_2} ( \hat{x}_2 ) ... Z_{m_N} ( \hat{x}_N ) \\
&= \sum_{{\bf{m}}\in \bf{S}} \prod_{l=1}^{N} Z_{m_l} ( \hat{x}_l ),
\label{NDZsum_vectors}
\end{split}
\end{equation}
with the sum including all $\bf{m}$ satisfying the resonance condition, ${\bf{q}} \cdot {\bf{m}} = p$. This expression enables an alternative route to deriving Eq.\ (\ref{HRWA}) starting from the power series defined in Eq.\ (\ref{Filter_RWA}). This route clarifies that single-resonance Hamiltonians include a product of 1D $Z$ functions, one per mode involved, while multi-resonance ones have a sum over terms of this form.

As discussed in the main text, the generating function can be used to give representations of these functions as integrals, Eq.\ (\ref{NDZint}).  Using this integral representation partial derivatives of the $Z$ functions with respect to any argument are found just to shift the index and bring down a factor $1/2$:

\begin{equation} \label{derivs}
\begin{split}
\frac{\partial}{\partial \hat{x}_j} Z_p^{\{ q \}} \left( {\bf{\hat{x}}} \right) &= \frac{1}{2}  Z_{p-q_j}^{\{ q \}} \left( {\bf{\hat{x}}} \right)
\\
\frac{\partial}{\partial \hat{x}_j^{\dagger}} Z_p^{\{ q \}} \left( {\bf{\hat{x}}} \right) &= -\frac{1}{2}  Z_{p+q_j}^{\{ q \}} \left( {\bf{\hat{x}}} \right)
\end{split} \, .
\end{equation}
These expressions are useful in deriving Eq.\ (\ref{fixed_point_equ}) and very useful in differentiating that expression with respect to each argument to determine the elements of the stability matrix.

Directly from the generating function, Eq. (\ref{generator}), one finds that reversing the sign of one of the superscript indices $q_j$ is equivalent to replacing the corresponding argument $\hat{x}_j$ by $-\hat{x}_j^\dagger$:
\begin{equation}
Z_{p}^{q_1,\hdots, -q_j, \hdots} \left( \bf{\hat{x}} \right) = Z_p^{ \{ q \} } \left( \hat{x}_1, \hdots, -{\hat{x}}_j^{\dagger},\hdots \right).
\end{equation}

Two more useful expressions can be derived from the integral from in Eq.\ (\ref{NDZint}) by manipulating the integration variable. First, by shifting the limits of the integral over $t$ and exploiting the periodicity one finds:
\begin{equation}
Z_p^{\{ q \}} \left( {\bf{\hat{x}}} \right) = e^{-\im p \theta} Z_p^{\{ q \}} \left( \hat{x}_1 e^{\im q_1 \theta},...,\hat{x}_N e^{\im q_N \theta}  \right).
\label{Phase_form}
\end{equation}
Second the periodicity can be used to see that multiplying all indices (both head and foot ones) by a single integer, $j$, leaves the expression unchanged:
\begin{equation}
Z_{j p}^{ j q_1, j q_2,\hdots,j q_N} \left( {\bf{\hat{x}}} \right) = Z_p^{q_1, q_2,..., q_N} \left( {\bf{\hat{x}}} \right).
\label{multiply_indicies_through}
\end{equation}

\bibliography{references}
\end{document}